\documentstyle[aps,epsf]{revtex}
\begin{document}
%@twocolumnfalse\endcsname

\author{ Osame Kinouchi\thanks{E-mail: osame@ultra3000.ifqsc.sc.usp.br }
Suani T. R. Pinho\thanks{E-mail:spinho@ gibbs.if.usp.br }  and
Carmem P. C. Prado\thanks{E-mail:prado@ if.usp.br }\\
Departamento de F\'{\i}sica Geral, Instituto de F\'{\i}sica \\
Universidade de S\~ao Paulo \\
Caixa Postal 66318,  CEP 05315-970 S\~ao Paulo, SP,  Brazil}

\title{On the random-neighbor Olami-Feder-Christensen slip-stick model}

\maketitle

\begin{abstract}
We reconsider the treatment of Lise and Jensen (Phys. Rev. Lett. {\bf 76},
2326 (1996)) on the random neighbor Olami-Feder-Christensen stik-slip model,
and examine the strong dependence of the results on the approximations used for
the distribution of states $p(E)$.

PACS number(s): 05.40.+j, 05.70.Jk, 05.70.Ln
 
\end{abstract} \bigskip

The work of Olami, Feder and Christensen \cite{OFC} on a slip-stick
earthquake model indicated, some time ago, that self-organized criticality
may occur without a local conservation law. Recently, it has been claimed by
Lise and Jensen \cite{LJ} that the random-neighbor version of the OFC model
also presents critical behavior above some critical dissipation level $%
\alpha _c<\alpha _0$, where $\alpha _0=1/q$ is associated with local
conservation for $q$ neighbors. These authors based their claims on some
theoretical mean field arguments and on numerical simulations with systems
with up to $N=400^2$\ sites. In order to perform the mean-field calculation they had
to make many different assumptions about the behavior of the model.

More recently, Chabanol and Hakin \cite{CH}, and Br\"oker and Grassberger 
\cite{BG} performed a more detailed analysis of the same model, showing that
what has been interpreted as a critical behavior in \cite{LJ} indeed
corresponds to a subcritical region with very large (but finite) mean
avalanche sizes. Although Br\"oker and Grassberger \cite{BG} gives a
comprehensive treatment of the random-neighbor version of the OFC model
(which we will designate R-OFC), it may be of interest to detect exactly
where the theoretical arguments given in \cite{LJ} fail, since\ that point
is not transparent in their paper and similar problems may occur or be of
interest in the future.\ This is the aim of our paper. We will show that the
problem is not in the method used in \cite{LJ} (which eventually can
give useful informations about the mechanism behind SOC) but in the strong
dependence of the output of the calculations on the exact form of  the
distribution of states $p(E)$ of the system.

To reinforce the strong dependence of the results on the specific form of $%
p(E)$, we revisit the R-OFC model, but this time introducing a simple and
small modification on the $p(E)$\ distribution, that consists in replacing
the interval $[0,E_c]$\ where the uniform distribution used by Lise and Jensen
was defined by the interval $[0,E^{\star }]$, with $E^{\star }<E_c$\ ($E_c$\
is the threshold value above which the sites become unstable and relax),
that is

\begin{equation}
\label{pe}p(E)=\frac 1{E^{\star }}\,\,\Theta (E)\,\,\Theta (E^{\star }-E),
\end{equation}
where $\Theta (x)$\ is the Heaviside function (see Figure (1-a)).

The random version of the OFC model (R-OFC) consists of  $N$ sites initially
with an energy $E_i<E_c$, for $i=1,...,N$. The sites with energy $E$ \
bellow $E_c$ are stable sites (inative) and will be labelled by the
superscript $-$; the sites with energy $E$ above $E_c$ are unstable (active)
and will be labelled by the superscript $+$. The energies of all sites are
increased slowly until the instant $t$ when the energy of a certain site 
$i$\ reaches the value $%
E_c$. This site becomes then unstable and the system relaxes in a very short
time scale according to the following rules:

\begin{eqnarray}
E_i(t+1) &=& 0 ,\nonumber\\ 
E_{rn}(t+1) &=& E_{rn}(t)+\alpha E^{+},
\end{eqnarray}

\noindent where $E_{rn}$\ stands for the energy of $q$\ other sites chosen at random, and $%
\alpha \leq 1/q$. Eventually some of these $q$\ sites may become unstable,
relax also, and so may generate an avalanching process that only stops when
the energies of all sites are again bellow $E_c$. If we have $\alpha =\alpha
_c=1/q$\ we say that the system is conservative. 

Following \cite{LJ}, the probability of an inactive site to be activated by
receiving a contribution $\alpha E^{+}$ of an active site is 

\begin{eqnarray}
P_+(E^+) &=& \frac{\int\limits_{E_c - \alpha E^+}^{E_c} p(E) 
dE}{\int\limits_{0}^{\infty} p(E) dE} \nonumber\\
&=& \frac{E^\star -E_c+\alpha E^+}{E^\star} .\
\end{eqnarray}

The branching ratio $\sigma $ is the average number of new unstable sites
created by a unstable site that relaxes. Clearly, in order to have a critical
branching process, we must have $\sigma =1$. For $q$ random-neighbors, we
have 

\begin{equation}
\sigma = q\;\frac{\int\limits_{E_c}^\infty dE^{+\,}P_{+}(E^{+})\,p(E^{+})}{%
\int\limits_{E_c}^\infty dE^{+}\,p(E^{+})} ,
\end{equation}

\noindent where $p(E^{+})$ is the distribution of states for unstable sites,\ that is,
sites with energy above $E^{\star}$. Adopting the notation $\langle
\ldots \rangle \equiv \frac{\int (\ldots )\,\,p(E)\,\,dE}{\int p(E)\,\,dE%
}$ for averages, we find 

\begin{equation}
\sigma = q\left[ \frac{E^{\star }-E_c}{E^{\star }}+\frac \alpha {E^{\star }}%
\langle E^{+}\rangle \right] .
\end{equation}

The critical branching ratio $\sigma =1$ defines a value $\alpha _c$ above
which infinite avalanches may occur.
The quantity $\langle E^{+}\rangle $ is estimated \cite{LJ} as 

\begin{equation}
\label{Emais}\langle E^{+}\rangle = \frac{\langle E^{-}\rangle }{1-\alpha }, 
\end{equation}

The average on $E^{-}$ is calculated as ,

\begin{eqnarray}
\label{eqe-}
\left\langle E^{-}\right\rangle \, &=& 
\frac{\int\limits_{E_c\,-\,\alpha
E^{+}}^{E_c}p(E)\,\,dE}{\int \limits_{E_c\,-\,\alpha
E^{+}}^{E_c}p(E)\,\,dE} \nonumber \\ 
&=& \frac{1}{E^{\star}-E_c +\alpha \,\left\langle 
E^{+}\right\rangle}\int\limits_{E_c\,-\,\alpha
\left\langle E^{+}\right\rangle }^{E^{\star }} E^- dE
\nonumber\\ 
&=& \frac{1}{2}\;\frac{\left( E^{\star}\right)^2-E_c^2+\left( 2E_c-\alpha
\,\left\langle E^{+}\right\rangle \right) \;\alpha \,\left\langle
E^{+}\right\rangle }{E^{\star }-E_c+\alpha \,\left\langle E^{+}\right\rangle }.
\end{eqnarray}

With this result and Eq.~(\ref{Emais}), we get

\begin{eqnarray}
(2\alpha -\alpha^2)x^2+\left[ 2E^{\star }(1-\alpha )-2E_c\right]
x+E_c^2-(E^{\star })^2 &=& 0,
\end{eqnarray}

where $x=\langle E^{+}\rangle $. This leads to the simple solution 

\begin{equation}
\langle E^{+}\rangle =\frac{E_c+E^{\star }}{2-\alpha }.
\end{equation}

Finally, we get for the branching ratio 

\begin{equation}
\sigma =q\left[ \frac{E^{\star }-E_c}{E^{\star }}+\frac{\alpha (E_c+E^{\star
})}{E^{\star }(2-\alpha )}\right] .
\end{equation}

The critical condition $\sigma =1$ leads to 

\begin{equation}
\alpha _c=\frac{E_c-E^{\star }(q-1)/q}{E_c+E^{\star }/(2q)}.
\end{equation}

For example,\ if  $q=4$ (case studied in \cite{LJ}), we have 

\begin{equation}
\alpha _c = \frac{E_c-3E^{\star }/4}{E_c+E^{\star }/8}.
\end{equation}

If we consider $E^{\star }=E_c$ we recover the Lise and Jensen value $\alpha
_c=2/9$. But the value of $\alpha _c$\ has a strong dependence on the value
of $E^{\star }$. We see that, already for the value $E^{\star
}/E_c=24/25=0.96$, $\alpha _c$ achieves the physical limit $0.25$.

In Figure 2 we show the behavior of the normalized coupling $\tilde 
\alpha _c=q\alpha _c/4$ as a function of $E^{\star }$. With this
normalization, the conservative case always corresponds to $\tilde \alpha
=0.25$ (as\ for $q=4$). We see that the allowed region of values for $%
E^{\star }$ so that $\tilde \alpha _c<0.25$ is very narrow for any $q$, and
that the value of $\alpha _c$ varies strongly in this region. We have
already shown in another work  (see \cite{PPK}) that already with lattices 
with $N=600^2$
sites it is possible to see a finite mean size avalanche for $\alpha =0.23$,
contrasting with the results of \cite{LJ} based in simulations in lattices
with $N=400^2$.

Besides showing that Lise and Jensen's approach is not robust with respect
to $p(E)$, we may ask about what kind of model produces the uniform
distribution 

\begin{equation}
\label{pcte}
p(E)=\Theta (E)\Theta \left( 1-E\right)  
\end{equation}

\noindent used by those authors. We found that an extremal version of the Feder and
Feder model, hereforth called EFF model, indeed produces this distribution.
Extremal here is used in the same sense it was first used in the Bak-Snappen
co-evolution model\cite{BS}. In the dynamics of extremal models 
there
is no driving step. We locate the site $i$ with the largest value of  $%
E_i=\max \left\{ E_j\right\}$ at the initial instant $t$. This site 
relaxes following the original FF rules:  

\begin{eqnarray}
E_i^m(t+1) &=& \eta ,\\ 
E_{nn}\left( t+1\right) &=& E_{nn}(t)+\alpha,\nonumber 
\end{eqnarray}

\noindent where $\eta $ is a noise, $\eta \in \left[ 0,\epsilon \right] $. If we
consider $\epsilon =\alpha =1/4$  we will have (\ref{pcte}). In this model the
size of an avalanche is defined as the number of sites with energy $E_i^m>1$ that
relaxes in a row.  Now, if we repeat the Lise and Jensen calculation using these EFF rules instead of the OFC rules, we get the self-consistent result  $\alpha_{c} = E_{c}/ q$.

It is also possible to show that a more realistic assumption about 
$p(E)$ leads to essentially the same results obtained by \cite{CH,BG}. If we
simulate the R-OFC model with $q=4$ we will get a energy distribution $p(E)$ with four peaks
\cite{LJ}.  They show clearly that $p(E)$ is not a simple constant. We then
decided to repeat the same calculations but supposing this time that \ $p(E)$\ had
the (more realistic) form shown in  figure (1-b), where $\Delta _p$ is 
half the
width of each peak and $\Delta _b$ is the width of the gaps between two
peaks. That means

\begin{equation}
p(E) = \left\{\begin{array}{ll} 
a, & \mbox{for $E\in I_{1}$ or $E\in I_{2}$ or $E\in I_{3}$ or $E\in 
I_{4}$}\\
0, & \mbox{otherwise}
\end{array}
\right.
\end{equation}

\noindent where $I_{1}= [0,\Delta_p]$, $I_{2}=[\Delta _p\!+\Delta _b,3\Delta _p+\Delta 
_b]$, $I_{3}=[3\Delta_p+\!2\Delta _b,5\Delta _p+2\Delta _b]$, and 
$I_{4}=[ 5\Delta _p+3\Delta _b,7\Delta _p+3\Delta _b]$. 

We also have that $E^{*}=3\Delta _b+7\Delta _p$ is the maximum value for which $p(E)\neq 0$.

\noindent Then we have

\begin{eqnarray}
P_{+}\left( E^{+}\right) &=& \frac{\int\limits_{E_c\,-\,\alpha
E^{+}}^{E_c}p(E)\,dE}{\int\limits_0^\infty p(E)\,dE} = 
\frac 1{7\,a\Delta _p}% \int\limits_{E_c\,-\,\alpha E^{+}}^{E_c}p(E)\,dE .
\end{eqnarray}

The lower limit of the integral in the numerator, $E_c-\alpha E^{+}$, can
now belong to any of the four intervals that define the peaks of the
distribution, to which we will asssign the indices $i=1,2,3,4.$ Considering
each one of the possibilities, the integrals $P_{+}^i\left( E^{+}\right) $
assume the generic form

\begin{equation}
P_{+}^i\left( E^{+}\right) =1+\frac{\left( i-1\right) \Delta _b}{7\Delta _p}-%
\frac{E_c}{7\Delta _p}+\frac{\alpha E^{+}}{7\Delta _p}. 
\end{equation}

The branching rate is given by  

\begin{equation}
\sigma = 4P_{+}^i = 4\left[ 1+\frac{\left( i-1\right) \Delta _b}{7\Delta 
_p}-%
\frac{E_c}{7\Delta _p}+\frac{\alpha \left\langle E^{+}\right\rangle }{%
7\Delta _p}\right].
\end{equation}

In a similar way used to obtain eq. (\ref{eqe-}), we calculate an 
expression for $\langle 
E^{-} \rangle$, which is associated with eq. (\ref{Emais}) to lead 

\begin{equation}
\label{Emaisi}
\left\langle E^{+}\right\rangle ^i = \frac{E_c}{\alpha \left( 2-\alpha 
\right) 
}-\frac{\left[ 7\Delta _p+\left( i-1\right) \Delta _b\right] \,\left(
1-\alpha \right) }{\alpha \left( 2-\alpha \right) }\pm \frac{\sqrt{\,y_i}}{%
2\alpha \left( 2-\alpha \right) },
\end{equation}

where 

\begin{equation}
y_i = 4\left\{ E_c\left( 1-\alpha \right) -\left[ 7\Delta _p+\left( 
i-1\right)
\Delta _b\right] \,\right\} ^2+4\alpha \left( 2-\alpha \right) \,\left[
x_i-14\left( i-1\right) \right] \,\Delta _p\,\Delta _b, 
\end{equation}

\noindent with $x_i=24,26,32$ and $42,$ for $i=1,2,3$ and $4$, respectively. Imposing
the branching condition $\sigma =1$ and using eq. (\ref{Emaisi}) we get%

\begin{equation}
7\Delta _p\left( 2+\alpha \right) +4\left( i-1\right) \Delta _b-4E_c\left(
1-\alpha \right) +7\alpha E_c\pm \sqrt{\,y_i} = 0 
\end{equation}

For instance, if we take $\Delta _p=0.08$ and $\Delta _b=0.1$ (that
corresponds to Figure (1-b)),  the critical branching condition leads to values
of $\alpha ^{*}$ outside the physical range (that is, $\alpha ^{\star}>1/4$).
Therefore, in this particular example, it is physically forbidden to assume
that $\sigma =1$, so there is no self-organized critical state.  

If we take the limit for the conservative case (that is, $\Delta
_p\rightarrow 0$ and $\Delta _b\rightarrow \alpha E_c$), the four peaks tend
to four delta functions at $(i-1)\,\alpha \,E_c$, and it is easy to see that
the condition $\sigma =1$  leads to the only possibility $\alpha ^{*}=\alpha
_c=1/4$ (we obtain $\alpha ^{*}>1/4$ for $i=1,2,3$). It can also be shown
that, if we consider the limit {\bf \ }$\Delta _b\rightarrow 0$\ and $\Delta
_p\rightarrow E_c\,/\,7$, (that is $p(E)$\ is constant in the interval $%
\left[ 0,E_c\right] $\ which corresponds to the approximation of ref. \cite
{LJ}) then $\alpha ^{*}=2/9$.

In general, for all values of $i,$ the regions of the parameter space
associated with $\alpha \leq 1/4$ are determined by%

\begin{equation}
E_c-\frac{175}{24}\Delta _p-\frac{2x_i}{21}\Delta _b\leq 0. 
\end{equation}

From this inequality plus the relation $E_c\geq 7\Delta _p+3\Delta _b$, (see
Figure 3) we see that only for a very small range  of the parameters 
$\Delta _p$
and $\Delta _b$ there are values of $\alpha _c$ in the physical range ($%
0<\alpha _c\leq 1/4$). In all of those cases, $\Delta _b$ is very small and
the shape of $p(E)$ is very close to the constant form used by Lise and Jensen.  Moreover, $\alpha$ varies strongly in these allowed ranges.

In conclusion, we showed that, besides not having considered lattices big
enough, the problem with   the approach used by Lise and Jensen in \cite{LJ}
is not in the method itself, but in the strong dependence of the output of
the calculations on the compatibility between the distribution $p(E)$ and the assumed dynamical rules which presumably lead to it. We also showed that the $p(E)$ approximation used by Lise and Jensen
does not correspond to the model they intend to analyse, namely R-OFC, but
to another non-conservative model, designated by us EFF (Extremal Feder and
Feder model). If we adopt the EFF's dynamical rules, the Lise and 
Jensen's method will lead to the right 
conclusions. In the end, we followed the same
approach but now considering a more realistic approach for  $p(E)$ and get
essentially the same results that had already been obtained through the use
of other arguments in \cite{CH,BG}, that is in the R-OFC model there is SOC
only in the conservative limit. 

Aknowledgments: We are thankful to S. R. Salinas for useful discussions 
about the analytical results of the R-OFC model. S. T. R. P. is on 
leave from Instituto de F\'{\i}sica, Universidade Federal da 
Bahia and acknowledges the support by the brazilian agency CAPES-PICD.

\pagebreak

\pagebreak

{\bf Figure Captions}

{\bf Fig. 1- a)} The uniform approximation for the energy distribution 
$p(E)$ of the R-OFC model. The solid line corresponds to the interval 
of $E$ used by Lise and Jensen, $[0,E_{c}]$; the dashed line corresponds 
to a smaller interval of $E$, $[0,E^{\star}]$, used in our calculations. {\bf b)} A more realistic 
approximation (non-uniform) of the energy distribution of the R-OFC 
model, for  $q=4$ (with four peaks). The parameters $\Delta_{p}$ and 
$\Delta_{b}$ are, respectively, the half-width of the peaks and the width 
of the gaps.  The value of $p(E)$ at the peaks is {\bf a} and $E_{c}\ge 
E^{\star}=7\Delta_{p}+3\Delta_{b}$.

{\bf Fig. 2-}  Normalized `critical' dissipation level 
$\tilde{\alpha}_c=k\alpha/4$ as a function of
$E^\star$. Values for $\tilde{\alpha}_c$ above $0.25$ 
are not physically admissible.

{\bf Fig. 3-} Space of parameters for $p(E)$ in terms of $\gamma _p=\Delta _p/E_c$
and $\gamma _b=\Delta _b/E_c$. The shaded regions correspond to the
intersection between $\alpha \leq 1/4$ and $\left( 7\Delta _p+3\Delta
_b\right) /E_c=7\gamma _p+3\gamma _b\leq 1$. Depending on the value of $%
E_c-\alpha E^{+}$, we have (a) $E_c-\alpha E^{+}$ $\epsilon $ $\left[
0,\Delta _p\right] $; (b) $E_c-\alpha E^{+}$ $\epsilon $ $\left[ \Delta
_p+\Delta _b,3\Delta _p+\Delta _b\right] $; (c) $E_c-\alpha E^{+}$ $\epsilon 
$ $\left[ 3\Delta _p+2\Delta _b,5\Delta _p+2\Delta _b\right] $; and (d) $%
E_c-\alpha E^{+}$ $\epsilon $ $\left[ 5\Delta _p+3\Delta _b,7\Delta
_p+3\Delta _b\right] $.
-

\begin{references}
\bibitem{OFC}  Z. Olami, H. J. F. Feder and K. Christensen, Phys. Rev. Lett. 
{\bf 68} 1244 (1992).

\bibitem{LJ}  S. Lise and H. J. Jensen,  Phys. Rev. Lett. {\bf 76}
2326 (1996).

\bibitem{CH}  M-L. Chabanol and V. Hakin, Phys. Rev. E {\bf 56} R2343
(1997).

\bibitem{BG}  H-M. Br\"oker and P. Grassberger, Phys. Rev. E {\bf 56} 3944
(1997).

\bibitem{PPK}  S. T. R. Pinho, C. P. C. Prado and O. Kinouchi, to be
published in Physica A, proccedings of the V Latin America Workshop on Non-linear Phenomena .

\bibitem{BS}  P. Bak and K. Sneppen, Phys. Rev. Lett {\bf71} 4083 (1993)
\end{references}
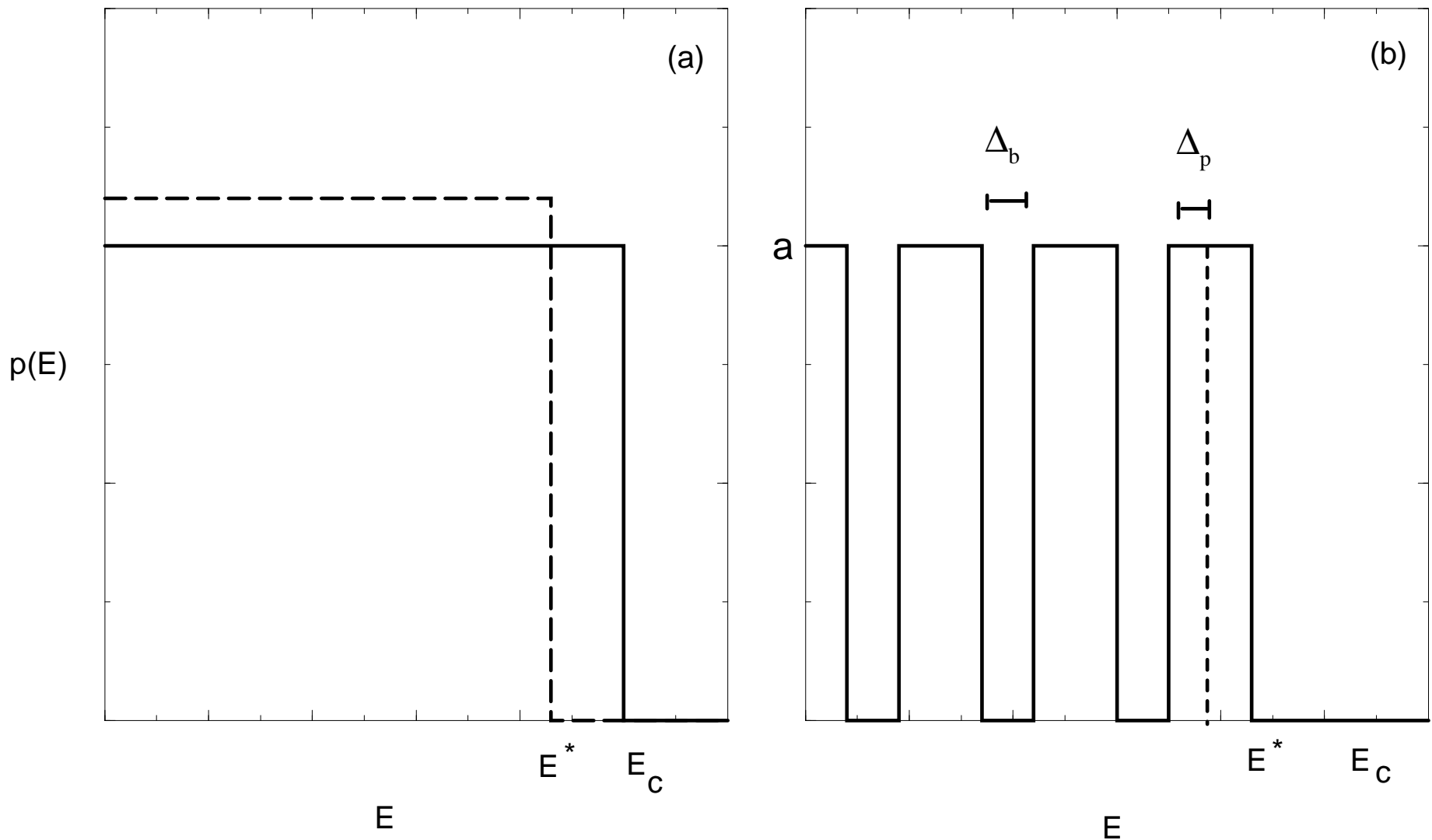
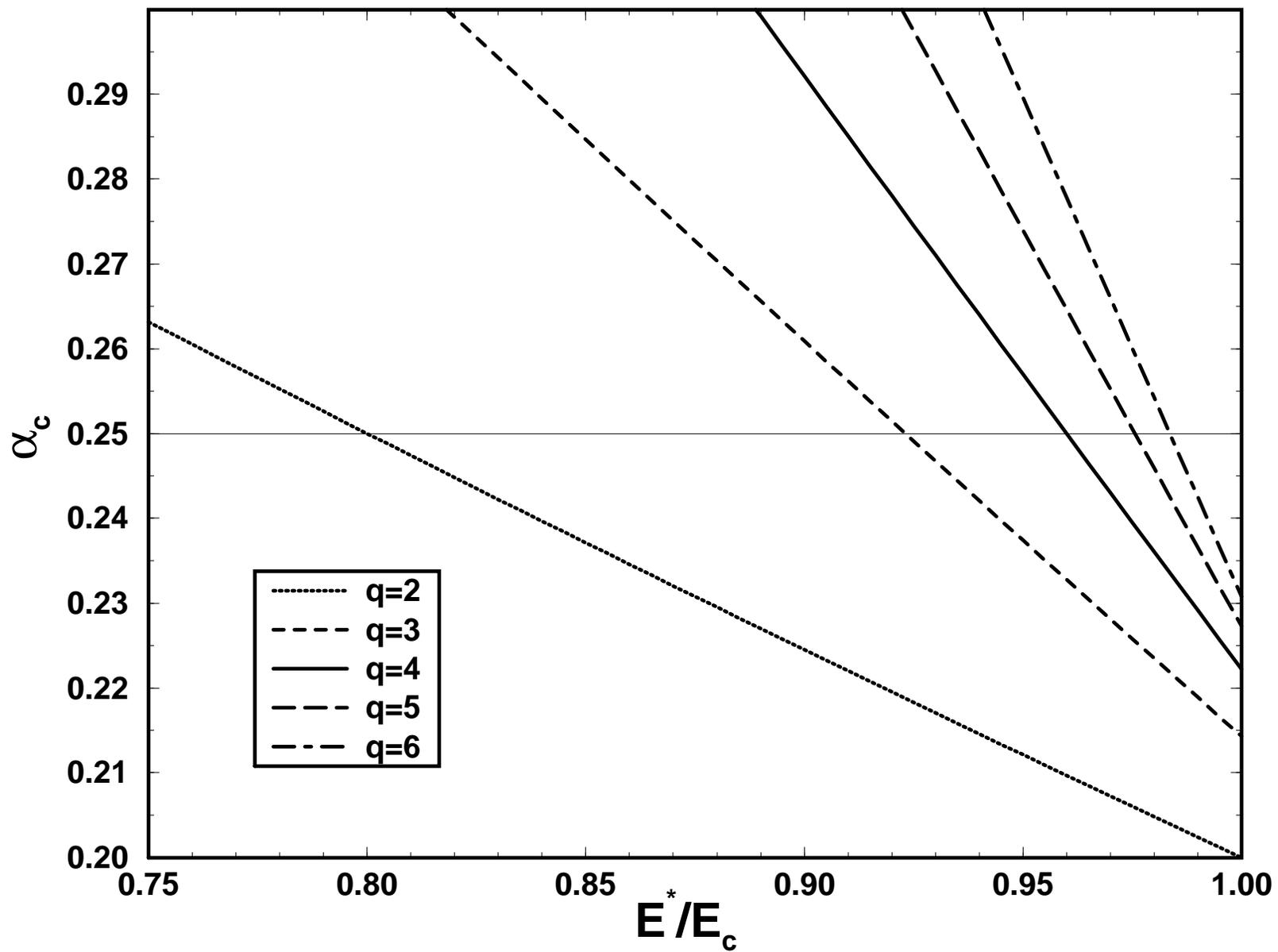
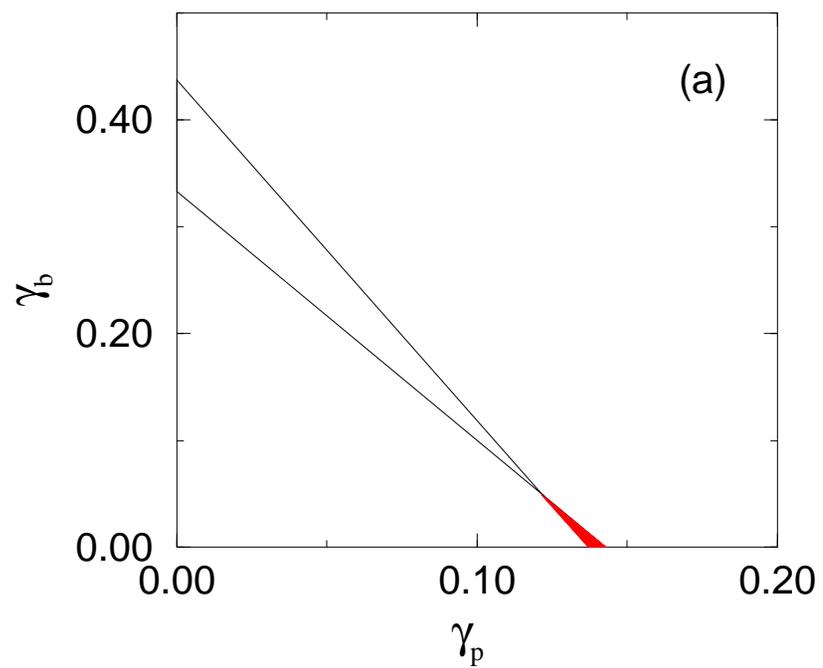
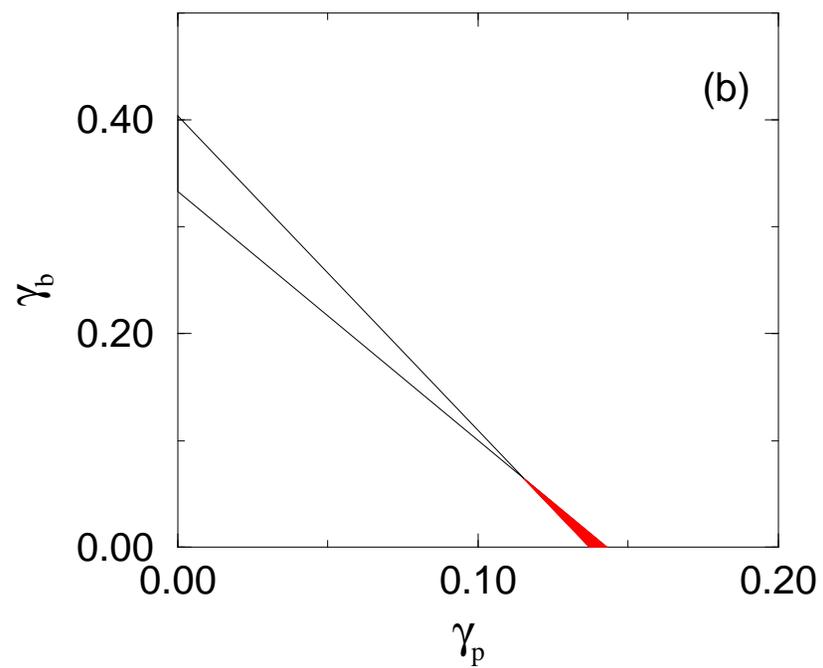
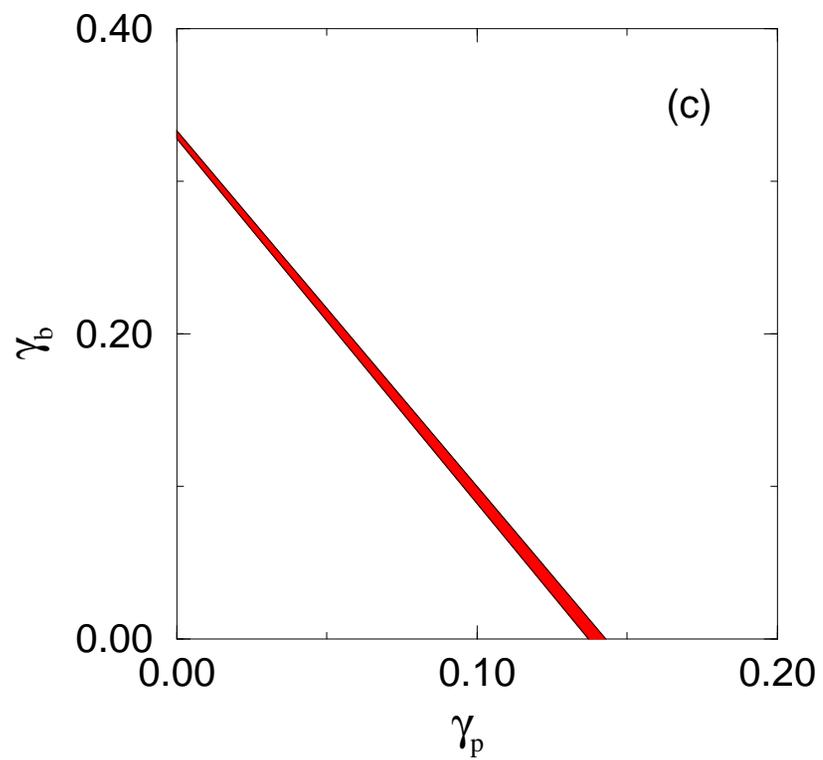
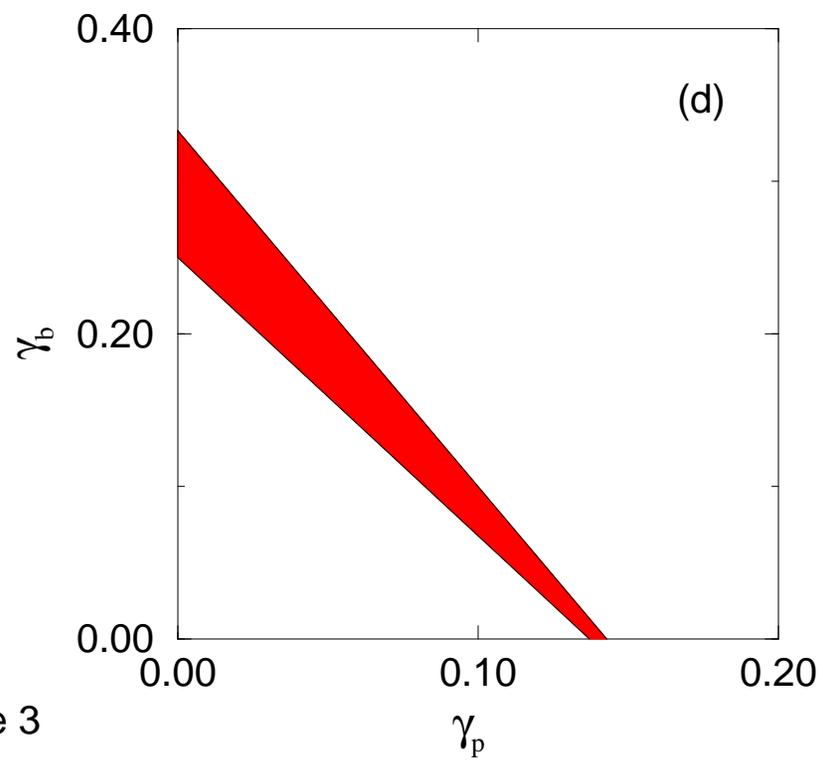
\end{document}